# Modelling Formation of Online Temporal Communities

Isa Inuwa-Dutse
Department of Computer Science Edge
Hill University, UK
dutsei@edgehill.ac.uk

## ABSTRACT

Contemporary social media networks can be viewed as a break to the early *two-step flow* model in which influential individuals act as intermediaries between the media and the public for information diffusion. Today's social media platforms enable users to both generate and consume online contents. Users continuously engage and disengage in discussions with varying degrees of interaction leading to formation of distinct online communities. Such communities are often formed at high-level either based on metadata, such as hashtags on Twitter, or popular content triggered by few influential users. These online communities often do not reflect true connectivity and lack the cohesiveness of traditional communities. In this study, we investigate real-time formation of temporal communities on Twitter. We aim at defining both high and low levels connections and to reveal the magnitude of clustering cohesion on temporal basis. Inspired by a real-life *event center* sitting arrangement scenario, the proposed method aims to cluster users into distinct and cohesive online temporal communities. Membership to a community relies on intrinsic tweet properties to define similarity as the basis for interaction networks. The proposed method can be useful for local event monitoring and clique-based marketing among other applications.

## KEYWORDS

Social networks mining; community formation; social network analysis; temporal communities; online social media; Twitter

## 1 INTRODUCTION

Users of modern day social media platforms such as Facebook and Twitter assume dual role as producers and consumers of huge amount of information.

Unlike the early two-step flow model of Katz and Lazarsfeld [1] in which influential users mediate communication, the influence network model of Watts and Dodds [2] enables multi-way flow of information where users can generate and consume information. The communication model of contemporary social media can be likened to the influence network model [2] by enabling multi-way flow of information. This property contrasts with the two-step flow model [1] that places some restrictions. However, the Twitter ecosystem can be seen to follow the *two-step* flow model based on apparent logical dichotomy: cliques of content pushers and consumers (Fig. 1).

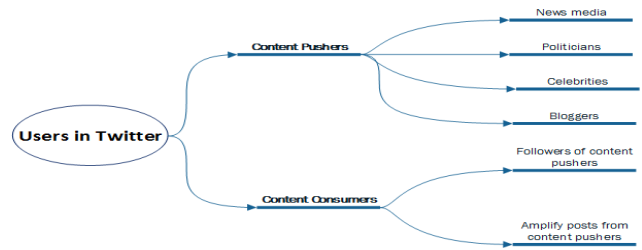

**Figure 1: Twitter users: Content pushers, such as news channels, influential bloggers, organisations, celebrities and politicians, often generate high volume of tweets and attract many followers. Content consumers, i.e. followers of such content pushers, further amplify their posts.**

Platforms such as Facebook and Twitter support continuous engagement and disengagement of diverse users leading to formation of online communities of varying degrees of cohesiveness. Some online communities can be easily discerned For example, groups on Facebook are formed explicitly and Twitter supports *followers-followees* and the use of hashtags [3]. Such communities are characterised as *explicit* [4]. Explicit communities are high-level and often not reflective of true connectivity [5]. Studies in this direction focus on high-level communities partly due to relying on data from popular hashtags or trending topics on Twitter which limit full realisation of benefits in communities such as cliquishness and local coherence. Moreover, while the continuous flow of tweets in real-time enables the formation of communities, studies tend to focus on static network data rather than real-time data. Implicit communities are dynamic and need exploratory activities to be discovered [4]. Such communities may undergo series of changes from birth, growth, contraction, merging, splitting, to death and concrete understanding of the underlying mechanisms governing this dynamism is still not fully understood [6].

We argue that despite the freedom and flexibility to disseminate information on current platforms, the flow of information is still influenced by few making it difficult to detect low level community of users interacting cohesively. In Fig 1, if we exclude *content pushers/influential users* (e.g., celebrities and news channels) from the networks, we will face a big layer of users which engage and form communities at microscopic level. This allows to consider connections of different granularity among users and to cluster users into distinct communities in real time. Users continuously form cohesive communities not based on trending topics or popular hashtags, without much attention from the literature. Consequently, we investigate real-time formation of temporal communities at microscopic level on Twitter using an algorithm inspired by a real-life event center sitting arrangement scenario to cluster users into distinct and cohesive communities.

Section II briefly presents background on online communities. Section III reviews related work. Section IV describes our community detection algorithm and Section V presents evaluation and preliminary results. Section VI concludes the study and proposes future work.

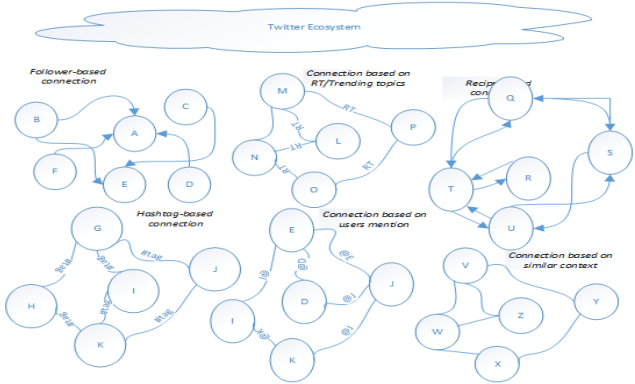

**Figure 2: Permanent or temporal pairwise connections between users based on the topic of discussion, retweeting, followership, friendship or use of similar hashtags on Twitter.**

## 2 BACKGROUND

### 2.1 Online Communities

Phenomena in real life are associated with numerous network structures and embedded communities. Classic models such as the random graph [7], small-world [8] and scale-free networks [9] form the basis on which networks and communities are studied. The social media ecosystem consists of numerous platforms with complex structures that enable interaction of diverse users. Twitter platform defines a low-level access to news in real-time distinguishing it from conventional news media that require extra effort such as editing and gadgets [3, 10]. Users ensure continuous flow of large volumes of tweets in real-time enabling the formation of communities of users at various level of granularity (Fig. 2). Users in the same community exhibit a high degree of similarity, while users in different communities exhibit no or low similarity [11, 12, 13]. Fig. 2 depicts how Twitter supports different granularity of connections among users. Users can be connected almost randomly: each user may use one or more hashtags to subscribe to a topic; user can also send direct messages and mention other users thereby creating a random connection. The figure also shows how message tags e.g. hashtags, can envelope related discussions leading to communities of users. However, Twitter's posting flexibility results in a variety of information making detection of cohesive groups difficult. The Twitter platform also acknowledges this challenge [14].

## 3 RELATED WORK

We review studies that share similar components, ideas and motivation with our approach. Related studies focus on clustering, topic/event detection [15], and information diffusion [16].

*Clustering and topic detection.* Clustering tasks are categorised as graph-based or similarity-based [13]. Graph-based approaches involve detecting subsets in graphs exhibiting dense intra-cluster and sparse inter-cluster connectivity [17]. Metrics, such as betweenness or shortest loop edges, are central in detection algorithms that process graphs to detect groups [18]. Pons and Latapy [19] developed an approach for hierarchical agglomerative clustering of objects based on the random walk model of Erdos and Renyi [7]. The approach assumes the existence of clusters in the network, which is not guaranteed in the case of real-time dynamic tweets. The sparsification technique in [20] relies on influential users as the basis for community formation. This leads to identifying followers as against cohesive user communities at low-level. The Louvain detection algorithm [21] uses the optimising modularity technique of Neumann [22]. The algorithm assumes the existence of community structure in the data. A variant of this algorithm was utilised to cluster a static collection of tweets disregarding the temporal effect [23]. Tsur [24] proposes a clustering technique based on message tags, such as hashtags, on Twitter. In [25] tweets are clustered by distinguishing posts related to real-world events from posts related to non-events. A method for automatic clustering and classification of tweets into subcategories on the basis of discussion hashtags is also proposed in [26]. These approaches detect high-level communities of users [27] with limited conformity between hashtags and associated sets of tweets. Although hashtags can be representative of the topics being discussed on Twitter [28], they are quite limited in exposing them fully. Users may use keywords relevant to the trending discussions but the content may not reflect the keyword and ultimately lead to communities with vague sense.

*Topic detection.* Topics on Twitter are associated with varying degrees of extent and intensity, with interesting topics exhibiting high activity burstiness [29]. An event detection method that relies on keywords used to define events on Twitter is proposed in [15]. A recent study analyses the relationship between users and discussion topics for community detection on Twitter [30]. The study, however, focuses on a single discussion topic per group. This is rarely the case, since users participate in many discussions.

*Time-based clustering.* The works of [10, 31 and 32] share our motivation in recognising the role of the temporal aspect in clustering dynamic network data. Similarly to community detection based on trending topics, Cataldi et al. [10] focuses on the popularity of terms over time. This approach is limited in capturing the full spectrum of communities especially when terms are becoming less popular. Another drawback is the retrospective comparison of terms within a definite time frame. Retrieving previous terms adds extra complexity in real-time deployment.

During its life-cycle on Twitter, the popularity of an average hashtag is uneven, exhibiting peaks and drops at various points in time. This property motivated the development of TOT-MMM, a system that captures the temporal clustering effect of latent topics in tweets [31]. As a hashtag-based approach, token similarity and other essential aspects of cohesive communities are overlooked. Similarly, the work of [32] relies on hashtags to cluster activity. Tweets not associated with hashtag are ignored, which ultimately leads to less cohesive communities because contextual properties are not fully represented.

Surveyed studies focus on high-level meta-data, such as hashtags, or static trending content with many users. Crucial parts of the network with no subscription to trending topics is concealed [33]. With an average of 6000 tweets per second [34], it is highly likely that low trending topics or events will be buried and communities of such users will remain undetected. Our goal is to address these limitations and aspire for low-level clustering of users by utilising intrinsic properties of tweets (Table 2) in real-time.

## 4 METHODOLOGY

The formation of temporal online user communities in real time on Twitter can be seen as a clustering task. Algorithms for partitioning a network into clusters fall under hierarchical or non-hierarchical methods [35]. For improved real-time efficiency, the proposed approach follows a non-hierarchical single-pass paradigm. Non-hierarchical methods assign objects to a pre-determined number of clusters incrementally [36, 37].

### 4.1 Problem Formulation

A community, $C$, is formed when a set of users, $U$ interacts based on textual content, $D$, within a given time period, $T$, resulting in distinct groups of users with common interests. As tweets are produced in real time (Fig. 3), the clustering process is similar to a real-life event center sitting arrangement.

Guests arrive in a venue or event and sit around tables similarly to how tweets are produce in real-time (Fig. 3). This intuition based inspired our approach:

*N guests ($g_1$ to $g_N$) attend an event, E. S tables, $s_1$ to $s_M$ (where m<N) are provided to accommodate a maximum of 6 guests each. Between (time $t_0 - t$), guest $g_1$ arrives first and sits on one of the tables ($s_i$). A few minutes later, $g_2$ arrives and sits on a table different from $g_1$. Guests $g_3$ and $g_4$ sit next to $g_1$ and $g_2$ respectively. Knowing $g_2$ for long, $g_5$ sits next to them. Guests arriving at different times (within $t_0 - t_1$) sit at a suitable table with space.* Fig. 4 illustrates this scenario.

In this scenario, small groups of guests are formed. Each table is a distinct and cohesive cluster of guests at a given time. We model this idea to capture how temporal communities on Twitter are formed and dispersed on a regular basis.

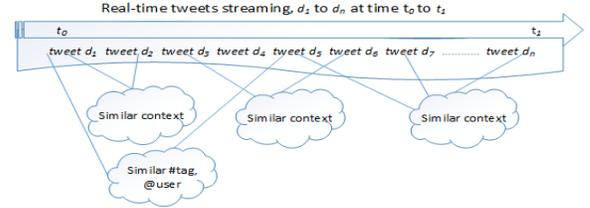

**Figure 3: Activity flow depicting real-time tweets generation and relevant fields to extract for cluster formation.**

Following the philosophy of incremental algorithm [36, 37], the clustering process in Fig. 4 starts with four random guests $g_a$, $g_b$, $g_c$, $g_d$ as seeds and compute similarities of each pair, e.g. ($g_a$, $g_b$) and ($g_c$, $g_d$) or ($g_a$, $g_c$) and ($g_b$, $g_d$), to initialise clusters. Dissimilar seeds form distinct clusters, e.g. $C_a$ and $C_b$, and similar seeds are assigned to the same cluster, e.g. $C_{ab}$. The process continues by assigning subsequent objects to existing clusters or forming new ones until the upper bound $M$ is reached. When the maximum number of clusters, $M$, is reached, a new batch of communities will be initiated. At the moment, M is an integer to be assigned heuristically according to the context.

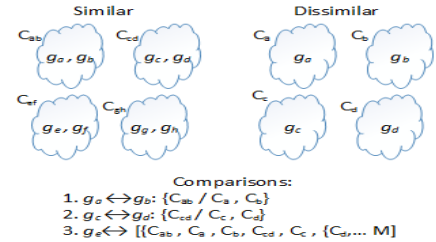

**Figure 4: Clusters formation and subsequent user assignment.**

Given a stream of tweets, $D$, generated at time $T$, as depicted in Fig. 3, $d_i t_i$ and $d_j t_j$ denotes a pair of tweets, $d_i$ and $d_j$, generated at times $t_i$ and $t_j$, respectively. The similarity function is given by:

$$sim(d_i, d_j) = \frac{||n_{d_i}|| \cup ||n_{d_j}||}{\sqrt{||n_{d_i}||} \times \sqrt{||n_{d_j}||}} \quad (1)$$

In Eqn. 1, $n_{d_i}$ and $n_{d_j}$ refer to the number of tokens in tweets $d_i$ and $d_j$. Geometrically, clusters $C = \{c_1, c_2, ..., c_m\}$ are assumed to be linearly independent, i.e. orthogonal to each other in the vector space. The goal is to minimise the angle between the centroid of a cluster $\mu$ and the data points in $D$ (Eqn. 2):

$$J(C_i) = \sum_{i=1}^{m} d_i - \mu_i \quad (2)$$

The function also minimises the angle between a data point $d_i$ and the corresponding cluster.

*Algorithm Pseudocode:*
For all users engaged in discussions on Twitter:
1. randomly pick a pair of users $d_i$ and $d_j$ at time $t$
2. compute their similarity, $sim(d_i, d_j)$
3. compare and assign users to clusters as follows:
    i. if similar, assign users to the new cluster $(d_i d_j)$

ii.    otherwise, assign them to new clusters $(d_i)$, $(d_j)$
    4. Repeat 1 - 3, while: number of clusters < M ($c_1$ to $c_M$)
        i.    Stop cluster formation
    5. Assign the remaining users to appropriate clusters:
        i.    Compute similarity of a user and clusters $c_1$ - $c_m$
        ii.   Assign user to the most similar cluster
    6. Stop

The implementation model consist of: an *input phase* to receive a stream of tweets *D* at discretised time *T*; a *clustering phase* for cluster formation and assignment of tweets using (a) the *ContSim function* to measure contextual similarity between tweets, (b) the *MetSim function* to measure similarity in metadata such as hashtags and user mentions and (c) the *AggSim function* to aggregate similarity as a function of *ContSim* and *MetSim;* the *output phase* to return cohesive list *l* of microscopic communities of users; and finally the *dispersion phase,* a time-based function to monitor dynamics of the formed communities and predict fading away or collapsing into another cluster.

## 5   PRELIMINARY RESULTS

As this is an on-going study, we present some preliminary results and relevant evaluation techniques. As a first step, we follow [38] in exploring the suitability of the data for clustering. The activity testbed utilised the non-hierarchical *k–means++* clustering algorithm in the *Scikit-Learn Toolkit* [39] to explore clustering tendency in the dataset.

### 5.1   Dataset

We use a dataset of 17,500 tweets collected in December, 2017 via the Twitter streaming API [40] using keywords related to the EU refugee crisis. The data was filtered heuristically to retain (1) users whose accounts are not verified by Twitter (2) tweets with relatively high number of hashtags and users mentions. We also ensure that the data is not from trending discussions. This ensures that popular contents will not dominate during clustering. Fig. 5 shows the relative proportions of the basic features and users.

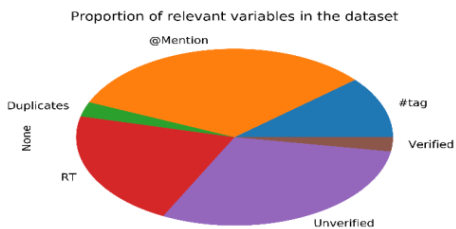

**Figure 5: Relative proportion of features and users in the data.**

### 5.2   Experimentation

The relatively small data size allows to fix the number of clusters and experiment using *k–means++*. Basic data preprocessing involved stopwords filtering and conversion to lowercase. TFIDF weighting was used to transform the data into numbers. The algorithm converged in 300 iterations and produced the clusters in Table 1. Fig. 6 shows a *2-d* visualization of common terms in some clusters in the output of the algorithm. For our future work, the clusters suggest that the data is suitable for clustering.

**Table 1: Example of clusters in the data**

| | |
|---|---|
| C0: | migrants European governments new abuse complicit com |
| C1: | victory misery launching phillipadams_1 tampa howard dar |
| C2: | china building along network camps border north |
| C3: | north koreans fears flood county refugee constructing |
| C4: | vous organizes calai harcelez dela postures entre |
| C5: | back Brexit urged mps caluse bill project |
| C6: | people notice communism always jackposobiec  flee |

*Interaction Graph.* Graph-based approaches have been proven to be effective in recognising explicit communities in networks of nodes and weighted edges [6]. In our case, explicit connectivity links between users may not be available. Thus, such approaches are unsuitable for real-time deployment. We utilise token-level similarity (Table 2) for graph formation. Fig. 7 depicts users as nodes. Edges are weighted by similarity scores.

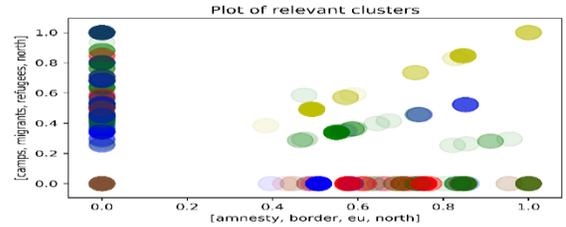

**Figure 6: Examples of common terms in the discovered clusters. The terms mostly relate to discussions about the EU refugee crisis.**

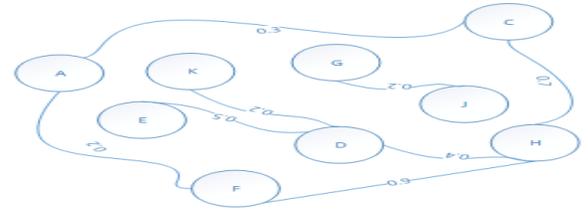

**Figure 7: Interaction graph of users with weighted edges (0–1) based on similarity magnitude. Higher values indicate stronger links.**

**Table 2: Similarity metrics to assess cohesiveness level**

| | *Cohesiveness* | Highly | High | Moderate | Low | Others |
|---|---|---|---|---|---|---|
| **Metric** | #Hashtag | √ | | √ | √ | |
| | @User | √ | | √ | √ | √ |
| | Follower | √ | √ | | √ | √ |
| | Followee | √ | √ | | | |
| | Friendship | √ | √ | | | |
| | Tweet | √ | √ | √ | | |

### 5.3   Evaluation

In explicit clustering, the quality of clusters can be evaluated using the silhouette coefficient [41] or the modularity technique of Neuman [42]. Non-hierarchical clustering requires a systematic evaluation of all possible clusters formed. Such requirements are often infeasible and heuristics are employed to determine sub-optimal clusters [6]. We use cohesion magnitude to assess clusters quality (Table 2) and adopt relevant methods from the literature.

Additionally, the following strategies from [6] are used in evaluating cohesiveness: (1) quantification and prediction of community's longevity based on basic quantities such as its size and age and (2) determining interaction intensity based on the ratio of intra-group and inter-group edges (i.e. $I = \frac{w_{ig}}{w_{og}}$).

# 6  CONCLUSIONS

The growing volume of real-time streams of data prompts the fundamental need for effective clustering. We propose a method to detect formation of temporal dense communities of users at low level on Twitter. The method is motivated by the formation of human cliques during real-life events that relate to the virtual Twitter environment. As this work is in progress, we demonstrated the direction to proceed and some preliminary results to assess the clustering potential in the data. A notable challenge in this regard is data sparsity, due to the large amount of unique tokens, especially when applying token-level similarity [31]. Future focus will consider: (1) how removing influential and non-influential users can improve the detection algorithm; (2) alternative means to detect and discard influential nodes when '*verifying users*' on Twitter is not an option and (3) assessing the quality of clusters in terms of homogeneity of members, deciding on the upper bound value for *M* and effective temporal effects to understand community dynamics. Understanding these aspects will allow full specification of algorithm parameters and apply on real data to test its efficacy and compare with baseline models.

## ACKNOWLEDGMENTS

My gratitude to my able supervisor Dr Ioannis Korkontzelos and members of my supervisory team Dr Mark Liptrott and Prof Franco Rizzuto. My thanks also to the anonymous reviewers.

## REFERENCES

[1]  Katz, E. and Lazarsfeld, P.F., 1966. *Personal Influence, The part played by people in the flow of mass communications*. Transaction Publishers.
[2]  Watts, D.J. and Dodds, P.S., 2007. Influentials, networks, and public opinion formation. *Journal of consumer research*, *34*(4), pp.441-458.
[3]  Kwak, H., Lee, C., Park, H. and Moon, S., 2010. What is Twitter, a social network or a news media?. In *Proceedings of the 19th international conference on World wide web* (pp. 591-600). ACM.
[4]  Papadopoulos, S., Kompatsiaris, Y., Vakali, A., Spyridonos, P. 2012. Community detection in social media, performance and application considerations. *Journal of Data Mining Knowledge Discovery*, *24*(3), pp.515-554.
[5]  Wilson, C., Boe, B., Sala, A., Puttaswamy, K.P. and Zhao, B.Y., 2009. User interactions in social networks and their implications. In *Proceedings of the 4th ACM European conference on Computer systems* (pp. 205-218). ACM.
[6]  Palla, G., Barabási, A.L. and Vicsek, T., 2007. Quantifying social group evolution. *Nature*, *446*(7136), p.664.
[7]  Erdos, P. and Rényi, A., 1960. On the evolution of random graphs. Publ. Math. Inst. Hung. Acad. Sci, 5(1), pp.17-60
[8]  Watts, D.J. and Strogatz, S.H., 1998. Collective dynamics of 'small-world'networks. *nature*, *393*(6684), p.440.
[9]  Barabási, A.L. and Albert, R., 1999. Emergence of scaling in random networks. science, 286(5439), pp.509-512
[10] Cataldi, M., Di Caro, L. and Schifanella, C., 2010. Emerging topic detection on twitter based on temporal and social terms evaluation. In *Proceedings of the tenth international workshop on multimedia data mining* (p. 4). ACM.
[11] Sundaram, H., Lin, Y.R., De Choudhury, M. and Kelliher, A., 2012. Understanding community dynamics in online social networks: a multidisciplinary review. *IEEE Signal Processing Magazine*, *29*(2), pp.33-40.
[12] Lancichinett A, Fortunato S and Kertesz J, Detecting the overlapping and hierarchical community structure in complex network, New Journal of Physics 11 (2009) 033015(18pp).
[13] Newman M. E. J, Detecting community structure in networks. The European Physical Journal B 38, 321 – 330 (2004).
[14] Twitter cofounder Evan Williams wants developers to come back, [online] available at http://uk.businessinsider.com/evan-williams-on-twitter-and-developers-2015-7?r=US&IR=T ; accessed 13/02/18.
[15] Ruchi, P. and Kamalakar, K., 2013. ET: Events from tweets. In Proc. 22nd Int. Conf. World Wide Web Comput., Rio de Janeiro, Brazil (pp. 613-620).
[16] C. Lagnier, L. Denoyer, E. Gaussier, P. Gallinari. 2013. Predicting Information Diffusion in Social Networks using Content and User's Profiles. ECIR 2013, Moscow, Russia, pp.74-85, Lecture Notes in Computer Science.
[17] Newman, M.E., 2002. The structure and function of networks. Computer Physics Communications, 147(1-2), pp.40-45.
[18] M. Fiedler, A. Pothon, H. Siman, and K. P, Liou, SIAM J Matrix Anal. Appl. 11, 430 (1990).
[19] Pons, P. and Latapy, M., 2006. Computing communities in large networks using random walks. J. Graph Algorithms Appl., 10(2), pp.191-218.
[20] Karthick, S., Shalinie, S.M., Kollengode, C. and Priya, S.M., 2014, December. A Sparsification Technique for Faster Hierarchical Community Detection in Social Networks. In *Eco-friendly Computing and Communication Systems (ICECCS), 2014 3rd International Conference on* (pp. 29-34). IEEE.
[21] Blondel, V.D., Guillaume, J.L., Lambiotte, R. and Lefebvre, E., 2008. Fast unfolding of communities in large networks. *Journal of statistical mechanics: theory and experiment*, *2008*(10), p.P10008.
[22] Newman, M.E., 2004. Fast algorithm for detecting community structure in networks. *Physical review E*, *69*(6), p.066133.
[23] Kim, Y.H., Seo, S., Ha, Y.H., Lim, S. and Yoon, Y., 2013. Two applications of clustering techniques to twitter: Community detection and issue extraction. *Discrete dynamics in nature and society*, *2013*.
[24] Tsur, O., Littman, A. and Rappoport, A., 2013. Efficient Clustering of Short Messages into General Domains. In *ICWSM*.
[25] Becker, H., Naaman, M. and Gravano, L., 2011. Beyond Trending Topics: Real-World Event Identification on Twitter. *Icwsm*, *11*(2011), pp.438-441.
[26] Rosa, K.D., Shah, R., Lin, B., Gershman, A. and Frederking, R., 2011. Topical clustering of tweets. *Proceedings of the ACM SIGIR: SWSM*.
[27] Vicient, C. and Moreno, A., 2014, August. Unsupervised semantic clustering of Twitter hashtags. In *ECAI* (pp. 1119-1120).
[28] Dann, S., 2010. Twitter content classification. *First Monday*, *15*(12).
[29] Kleinberg, J., 2003. Bursty and hierarchical structure in streams. *Data Mining and Knowledge Discovery*, *7*(4), pp.373-397.
[30] Gadek, G., Pauchet, A., Malandain, N., Khelif, K., Vercouter, L. and Brunessaux, S., 2017. Topical cohesion of communities on Twitter. *Procedia Computer Science*, *112*, pp.584-593.
[31] Lu, H.M. and Lee, C.H., 2015. A twitter hashtag recommendation model that accommodates for temporal clustering effects. *IEEE Intelligent Systems*, *30*(3), pp.18-25.
[32] Feng, W., Zhang, C., Zhang, W., Han, J., Wang, J., Aggarwal, C. and Huang, J., 2015. STREAMCUBE: hierarchical spatio-temporal hashtag clustering for event exploration over the twitter stream. In *Data Engineering (ICDE), 2015 IEEE 31st International Conference on* (pp. 1561-1572). IEEE.
[33] Guille, A. and Favre, C., 2015. Event detection, tracking, and visualization in twitter: a mention-anomaly-based approach. *Social Network Analysis and Mining*, *5*(1), p.18.
[34] Social Media Statistics and Facts, [Online], available at https://www.statista.com/topics/1164/social-networks/ , accessed 13-02-2018.
[35] Manning, C.D. and Schütze, H., 1999. *Foundations of statistical natural language processing*. MIT press.
[36] Allan, J., Carbonell, J.G., Doddington, G., Yamron, J. and Yang, Y., 1998. Topic detection and tracking pilot study final report.
[37] Charikar, M., Chekuri, C., Feder, T. and Motwani, R., 2004. Incremental clustering and dynamic information retrieval. *SIAM Journal on Computing*, *33*(6), pp.1417-1440.
[38] Jain, A.K. and Dubes, R.C., 1988. Algorithms for clustering data.
[39] Pedregosa, F., Varoquaux, G., Gramfort, A., Michel, V., Thirion, B., Grisel, O., Blondel, M., Prettenhofer, P., Weiss, R., Dubourg, V. and Vanderplas, J., 2011. Scikit-learn: Machine learning in Python. *Journal of machine learning research*, *12*(Oct), pp.2825-2830.
[40] Twitter Streaming APIs, [Online] available at https://developer.twitter.com/, accessed 03-01-2018
[41] Kaufman, L. and Rousseeuw, P.J., 1990. Partitioning around medoids (program pam). *Finding groups in data: an introduction to cluster analysis*, pp.68-125.
[42] Newman, M.E., 2006. Modularity and community structure in networks. *Proceedings of the national academy of sciences*, *103*(23), pp.8577-8582.